\documentclass[12pt]{article}
\usepackage{graphicx}
\usepackage{amsmath}
\usepackage{amssymb}
\usepackage{caption2}
\begin{document}
\bibliographystyle{prsty}
\begin{center}
{\large {\bf \sc{  Magnetic moment of the pentaquark $\Theta^+(1540)$ as diquark-diquark-antiquark state
with QCD sum rules }}} \\[2mm]
Z. G. Wang$^{1}$ \footnote{Corresponding author;
E-mail,wangzgyiti@yahoo.com.cn.  }, S. L. Wan$^{2}$ and
W. M. Yang$^{2} $    \\
$^{1}$ Department of Physics, North China Electric Power University, Baoding 071003, P. R. China \\
$^{2}$ Department of Modern Physics, University of Science and Technology of China, Hefei 230026, P. R. China \\
\end{center}

\begin{abstract}
In this article, we study the magnetic moment of the pentaquark
state $ \Theta^+(1540)$ as diquark-diquark-antiquark
($[ud][ud]\bar{s}$) state with the QCD sum rules in the external
weak electromagnetic field (EFSR) and the light-cone QCD sum rules
(LCSR) respectively. The numerical results indicate the magnetic
moment is about $\mu_{\Theta^+}=-(0.11\pm 0.02)\mu_N$ for the EFSR
and $\mu_{\Theta^+}\approx-(0.1-0.5)\mu_N$ for the LCSR. As the
values obtained from the EFSR are more stable than the corresponding
ones from the LCSR, $\mu_{\Theta^+}=-(0.11\pm 0.02)\mu_N$ is more
reliable.
\end{abstract}

PACS : 12.38.Aw, 12.38.Lg, 12.39.Ba, 12.39.-x

{\bf{Key Words:}} QCD Sum Rules, Magnetic moment,  Pentaquark
\section{Introduction}

The observation of the new baryon state $\Theta^+(1540)$ with
positive strangeness and minimal quark content $udud\bar{s}$
\cite{exp2003} has motivated intense theoretical investigations
 to clarify the quantum numbers and to understand the
under-structures of the exotic state
\cite{DiakonovPentaquark,ReviewPenta}. Although the pentaquark state
$\Theta^+(1540)$ can be signed to the top of the antidecuplet
$\overline{10}$ with isospin $I=0$, the spin and parity have not
been experimentally determined yet and no consensus has ever been
reached  on the theoretical side
\cite{DiakonovPentaquark,ReviewPenta}. The discovery has opened a
new field of strong interaction  and provides a new opportunity for
a deeper understanding of the low energy QCD especially when
multiquark states are involved. The magnetic moments of the
pentaquark states are
  fundamental parameters as their  masses, which have  copious  information  about the underlying quark
structures, can be used to distinguish the preferred
configurations from  various  theoretical models and  deepen our
understanding of the underlying dynamics.
 Furthermore, the magnetic moment of the pentaquark state $\Theta^+(1540)$ is an important ingredient
in studying the cross sections of the photo- or
electro-production, which can be used to determine the fundamental
quantum number of the pentaquark state $\Theta^+(1540)$, such
  as spin and parity \cite{HosakaPP,Nam04M}, and may be extracted from experiments eventually in the
future.

 There have been several works on the magnetic moments of the pentaquark state
  $\Theta^+(1540)$ \cite{Huang04M,Zhao04M,Kim04M,Nam04M,Zhu04M,Inoue04M,Bijker04M,Goeke04M,Hong04M,Delgado04M,Wang05M},
in this article, we take the point of view that the baryon
$\Theta^{+}(1540)$ is a diquark-diquark-antiquark
($[ud][ud]\bar{s}$) state with the quantum numbers $J=\frac{1}{2}$ ,
$I=0$ , $S=+1$, and study its magnetic moment with the QCD sum rules
in the external weak electromagnetic field (EFSR) and the light-cone
QCD sum rules (LCSR) respectively
\cite{Shifman79,Ioffe84,LightCone}. Different quark configurations
can be implemented with different interpolating currents, if the $u$
and $d$ quarks in the pentaquark state $\Theta^+(1540)$ are bound
into spin zero, color and flavor antitriplet $\bar{3}$ diquarks, we
can take the diquarks (for example, $\epsilon^{abc} u_b C\gamma_5
d_c$ with $J^P=0^+$ and $\epsilon^{abc} u_b C d_c$ with $J^P=0^-$ )
instead of the $u$ and $d$ quarks as the basic constituents to
construct the interpolating currents.

 The article is arranged as follows:   we derive the EFSR and LCSR for the magnetic moment of the
pentaquark state $\Theta^+(1540)$ in section II; in section III,
numerical results and discussions; section IV is reserved for
conclusion.

\section{EFSR and LCSR}
 Although for medium and asymptotic momentum transfers
 the  operator product expansion approach  can be applied for the form factors and moments of wave
 functions, at low momentum transfer,
 the standard operator product expansion approach cannot be  consistently  applied,
 as pointed out in the early work on photon
 couplings at low momentum for the nucleon  magnetic moments
 \cite{Ioffe84}. In Ref.\cite{Ioffe84}, the problem was solved by using a
 two-point correlation function in an external electromagnetic field, with
  vacuum susceptibilities introduced as parameters for
 nonperturbative propagation in the external field, i.e. the QCD  sum rules in the external field.
 As  nonperturbative vacuum properties, the susceptibilities can be
 introduced for both small and large momentum transfers  in the
 external fields. The alternative way
 is the light-cone QCD sum rules, which
  was firstly used to calculate the magnetic moments of the
nucleons in Ref.\cite{Braun89}. For more discussions about the
magnetic moments of the baryons in the framework of the LCSR
approach, one can consult Ref.\cite{Aliev}.

In the following, we write down  the two-point correlation functions
$\Pi_{EF\eta}(p)$ and $\Pi_{LC\eta}(p)$ for the EFSR and LCSR
respectively \cite{Matheus04},
\begin{eqnarray}
\Pi_{EF\eta}(p)&=& i\int d^4 x e^{ip\cdot x}\langle 0|T \{
\eta(x){\bar \eta}(0) \}|0\rangle_{F_{\mu\nu
} } \, , \nonumber \\
&=&\Pi_0(p) + \Pi_{\mu\nu} (p) F^{\mu\nu} +\cdots \, , \\
 \Pi_{LC\eta} (p, q)&=&  i\int d^4 x \, e^{i p x}
\langle\gamma(q)|T\{\eta(x) \bar \eta(0) \}|0\rangle ,
\end{eqnarray}
where
\begin{eqnarray}
\eta_1(x)&=&{1\over\sqrt{2}}\epsilon^{abc}\left\{\left[u_a^T(x) C
\gamma_5 d_b(x)\right] \left[u_c^T(x) C \gamma_5
d_e(x)\right]C\bar{s}^T_e(x) -
(u\leftrightarrow d)\right\} , \\
\eta_2(x)&=&{1\over\sqrt{2}}\epsilon^{abc}\left\{\left[u_a^T(x) C
d_b(x)\right] \left[u_c^T(x) C  d_e(x) \right ]C\bar{s}^T_e(x) -
(u\leftrightarrow
d)\right\}  , \\
 \eta(x)&=&\left\{t \eta_1(x) + \eta_2(x) \right\} .
\end{eqnarray}
Here the $\gamma(q)$ represents the external electromagnetic field
 $A_\mu (x)= \varepsilon_\mu e^{iq\cdot x}
$, the $\varepsilon_\mu$ is the photon polarization vector and the
field strength  $F_{\mu\nu}(x)=i( \varepsilon_\nu q_\mu
-\varepsilon_\mu q_\nu) e^{iq\cdot x}$. The $\Pi_0(p)$ is the
correlation function  without the external field $F_{\mu\nu}$ and
the $\Pi_{\mu\nu}(p)$ is the linear response term. The $a$, $b$, $c$
and $e$ are color indexes,  the $C=-C^T$ is the charge conjugation
operator, and the $t$ is an arbitrary parameter. The constituents $
\epsilon^{abc} u_b^T(x)C\gamma_5 d_c(x) $ represent the scalar
diquarks with $J^P=0^+$ and $ \epsilon^{abc} u_b^T(x)Cd_c(x)  $
represent the pseudoscalar diquarks  with $J^P=0^-$, we can denote
the $\eta_1(x)$ and $\eta_2(x)$ by S-type and P-type interpolating
current respectively according to the spin and parity of the
constituent diquarks. They both belong to the antitriplet $\bar{3}$
representation of the  color and flavor $SU(3)$ group,  and can
cluster together with diquark-diquark-antiquark  structure to give
the total spin and parity for the pentaquark state $\Theta^+(1540)$
$J^P={\frac{1}{2}}^+$ \footnote{We can write down the interpolating
currents for the other pentaquark states in the multiplets
$\overline{10}+8$ based on the Jaffe-Wilczek's diquark model in the
same way as we have done in Eqs.(3-5), then preform the operator
product expansion and take the current-hadron duality to obtain the
magnetic moments. Comparing with the magnetic moments in the
multiplets and detailed studies may shed light on the
under-structures and low energy dynamics of the pentaquark states.
However, the calculations of the operator product expansion for  a
number of correlation functions are tedious and beyond the present
work, this may be our next work.}. The scalar diquarks correspond to
the $^1S_0$ states of $ud$ quark system. The one-gluon exchange
force and the instanton induced force can lead to significant
attractions between the quarks in the $0^+$ channels
\cite{GluonInstanton}. The pseudoscalar diquarks do not have
nonrelativistic limit,  can be taken as  the $^3P_0$ states.

At the level of hadronic degrees of freedom, the linear response
term $\Pi_{\mu\nu}(p)$ can be written as
\begin{equation}
\Pi_{\mu\nu}(p)F^{\mu\nu} = i\int d^4x  e^{ip\cdot x}\langle 0\,|
\eta(x) \left\{ -i \int d^4y A_\mu(y) J^\mu(y)\right\}
\bar{\eta}(0)\,|\,0\rangle.
\end{equation}
According to the basic assumption of current-hadron duality in the
QCD sum rules  approach \cite{Shifman79}, we insert  a complete
series of intermediate states satisfying the unitarity   principle
with the same quantum numbers as the current operator $\eta(x)$
 into the correlation functions in
Eq.(2) and Eq.(6)  to obtain the hadronic representation. After
isolating the double-pole terms of the lowest pentaquark  states, we
get the following results,
\begin{eqnarray}
\Pi_{\mu\nu }(p)F^{\mu\nu} & = & -\int d^4x \int d^4y {d^4 k \over
(2\pi)^4} {d^4 k^\prime \over (2\pi)^4} \sum_{ ss^\prime} \frac{1}
{m^2_{\Theta^+} -k^2-i\epsilon}\frac{1} {m^2_{\Theta^+}
-{k^\prime}^2-i\epsilon} \nonumber \\ & & e^{ip\cdot x} A_\mu(y)
\langle 0 | \eta(x) | ks \rangle \langle ks | J^\mu(y) | k^\prime
s^\prime \rangle \langle k^\prime s^\prime | \bar{\eta}(0) | 0
\rangle+\cdots\, \nonumber\\
    & = & -F^{\mu\nu}f_0^2 \frac{F_1(0)+F_2(0)}{4(m_{\Theta^+}^2-p^2)^2} \{ {\hat p}
      \sigma_{\mu\nu} +\sigma_{\mu\nu} {\hat p} \}+\cdots ;
\end{eqnarray}
\begin{eqnarray}
\Pi_{LC\eta} (p, q) &=& -f_0^2\varepsilon^\mu\frac{{\hat
p}+m_{\Theta^+}}{p^2-m_{\Theta^+}^2} [F_1(q^2)\gamma_\mu
+\frac{i\sigma_{\mu\nu}q^\nu }{2m_{\Theta^+}}F_2(q^2)]\frac{{\hat p}+{\hat q}+m_{\Theta^+}}{(p+q)^2-m_{\Theta^+}^2} +\cdots  \nonumber \\
 &=&-
\frac{f_0^2\left[F_1(q^2)+
F_2(q^2)\right]}{(p^2-m_{\Theta^+}^2)((p+q)^2-m_{\Theta^+}^2)} {\hat
p}
{\hat \varepsilon} ({\hat p}+{\hat q}) +\cdots \nonumber\\
&=&\Pi_{LC} (p,
q)i\varepsilon_{\mu\nu\alpha\beta}\gamma_5\gamma^\mu\varepsilon^\nu
q^\alpha p^\beta+\cdots.
\end{eqnarray}
Here we  have used  the fix-point gauge  $x_\mu A^\mu(x)=0$,
$A_\mu(y)=-{1\over 2} F_{\mu\nu} y^\nu $, and the definition
$\langle 0| \eta(0) |\Theta^+ (p)\rangle =f_0 u(p)$. From the
electromagnetic form factors $F_1(q^2)$ and $ F_2(q^2)$ , we can
obtain the
 magnetic moment of the pentaquark state $\Theta^+(1540)$,
\begin{equation}
\mu_{\Theta^+}=\left\{ F_1(0)+F_2(0)\right\}
\frac{e_{\Theta^+}}{2m_{\Theta^+}}.
\end{equation}
 The linear response term $\Pi_{\mu\nu}(p)$ in the weak
external electromagnetic field $F_{\mu\nu}$  has three different
Dirac tensor structures,
\begin{equation}
\Pi_{\mu\nu}(p) =\Pi_{EF} (p) \left\{\sigma_{\mu\nu} {\hat p} +{\hat
p}\sigma_{\mu\nu}\right\} +\Pi_1(p)
i\left\{p_{\mu}\gamma_{\nu}-p_{\nu}\gamma_{\mu}\right\}{\hat p}
+\Pi_2(p) \sigma_{\mu\nu}  \, .
\end{equation}
The first structure has an odd number of $\gamma$-matrix and
conserves chirality, the second and third have even number of
$\gamma$-matrixes and violate chirality. In the original QCD sum
rules analysis of the nucleon magnetic
 moments \cite{Ioffe84}, the interval of dimensions (of the condensates) for the odd
structure is larger than the interval of dimensions for the even
structures, one may expect a better accuracy of the results obtained
from the sum rules with  the odd structure. In this article, the
spin of the pentaquark state $\Theta^+(1540)$ is supposed to be
$\frac{1}{2}$, just like the nucleon.  As in  our previous work
\cite{Wang05M},  we can choose  the first Dirac tensor structure
$\left\{\sigma_{\mu\nu} {\hat p} +{\hat p}\sigma_{\mu\nu}\right\}$
for analysis. The phenomenological spectral density of the EFSR in
Eq.(7) can written as,
\begin{equation}
\frac{\mbox{Im} \Pi_{EF} (s)}{\pi} = {1\over 4} \{F_1(0)+F_2(0)\}
f_0^2 \delta^\prime (s-m_{\Theta^+}^2) + C_{subtract} \delta
(s-m_{\Theta^+}^2) +\cdots\, ,
\end{equation}
where the first term corresponds to the magnetic moment of the
pentaquark state $\Theta^+(1540)$,  and is of double-pole. The
second term comes from the electromagnetic transitions between the
pentaquark state $\Theta^+(1540)$
 and the excited states (or high resonances),  and is of single-pole. Here we introduce the quantity  $C_{subtract}$ to represent
 the electromagnetic transitions between the ground pentaquark state and the high
 resonances, it may have   complex dependence  on the energy $s$ and high
 resonance masses. However, we have no knowledge about the high resonances,
 even the existence  of the ground pentaquark state $\Theta^+(1540)$
 is not firmly established, which is in contrast to the conventional baryons, in those channels we can use
 the experimental data as a guide in constructing the phenomenological spectral densities
 \cite{Nielsen05}. In practical manipulations, we can take the $C_{subtract}$ as
 an  unknown constant, and fitted to reproduce reliable values for
 the form factors $F_1(0)+F_2(0)$, we will revisit this subject
 in Eq.(19).
  The contributions from the higher
resonances and  continuum states  are suppressed after Borel
transform  and not shown explicitly for simplicity.  For the LCSR,
  we write down only the double-pole term explicitly in Eq.(8) which corresponds to
the magnetic moment of the pentaquark state $\Theta^+(1540)$, and
choose  the tensor structure
$\varepsilon_{\mu\nu\alpha\beta}\gamma_5\gamma^\mu\varepsilon^\nu
q^\alpha p^\beta$ for analysis \cite{Huang04M}. The contributions
from the single-pole terms which concerning the excited and
continuum states are suppressed after the double Borel transform,
and not shown explicitly for simplicity.

 The  calculation of the operator product expansion in the  deep Euclidean space-time region at the level of
quark and gluon degrees of freedom is
  straightforward and tedious, here technical details are neglected for simplicity,
  once  the analytical  results are obtained,
  then we can express the correlation functions at the level of quark-gluon
degrees of freedom into the following forms through dispersion
relation,
  \begin{eqnarray}
  \Pi_{EF}(P^2)&=& \frac{e_s}{\pi}\int_{m_s^2}^{s_0}ds
  \frac{{\rm Im}[A(s)]}{s+P^2}+
  e_sB(P^2)+\cdots\, , \\
\Pi_{LC}(p,q)&=& e_s\int_0^1 du
\left\{\frac{1}{\pi}\int_{m_s^2}^{s_0}ds
  \frac{{\rm Im}[C(s)]}{s-up^2-(1-u)(p+q)^2}+
  D(p,q) \right\}+\cdots\, ,
  \end{eqnarray}
where
\begin{eqnarray}
\frac{{\rm Im}[A(s)]}{\pi}&=&
-\frac{(5t^2+2t+5)s^4}{2^{14}5!4!\pi^8}+\frac{(5t^2+2t+5)s^3m_s\chi
\langle \bar{s}s\rangle}{2^{9}5!4!\pi^6} +\frac{(5+2t-7t^2)s\langle
\bar{q}q\rangle^2}{2^{9}3^2\pi^4}\nonumber \\
&&-\frac{(5+2t-7t^2)m_s \langle \bar{q}q\rangle^2 \langle
\bar{s}s\rangle
\chi}{2^{7}3^2\pi^2}-\frac{(5t^2+2t+5)s^2}{2^{17}4!\pi^6}\langle
\frac{\alpha_s GG}{\pi} \rangle , \nonumber \\
B(P^2)&=&\frac{(5t^2+2t+5)\langle \bar{q}q\rangle^4}{2^{5}3^3 P^4},
\nonumber \\
 \frac{{\rm Im}[C(s)]}{\pi}&=&
\frac{(5t^2+2t+5)s^4}{2^{12}5!4!\pi^8}-\frac{(5t^2+2t+5)s^3f
\psi(u)}{2^{13}5!\pi^6}+\frac{(7t^2-2t-5)s\langle
\bar{q}q\rangle^2}{2^{7}3^2\pi^4}\nonumber \\
&&-\frac{(7t^2-2t-5)\langle \bar{q}q\rangle^2f
\psi(u)}{2^{6}3\pi^2}+\frac{(5t^2+2t+5)s^2}{2^{15}4!\pi^6}\langle
\frac{\alpha_s GG}{\pi} \rangle  \nonumber \\
&&-\frac{(5t^2+2t+5)sf \psi(u)}{2^{14}3\pi^4}\langle
\frac{\alpha_s GG}{\pi} \rangle ,\nonumber \\
D(p,q)&=&-\frac{(5t^2+2t+5)\langle \bar{q}q\rangle^4}{2^{3}3^3
\left(-up^2-(1-u)(p+q)^2\right)^2}. \nonumber
\end{eqnarray}
From Eqs.(12-13), we can see that due to the special interpolating
current $\eta(x)$ ( see Eqs.(3-5)), the $u$ and $d$ quarks which
constitute the diquarks have no contributions to the magnetic moment
 though they have electromagnetic interactions with the external
field, the net contributions to the magnetic moment come from the
$s$ quark only, which is  different significantly from the results
obtained in Refs.\cite{Huang04M,Wang05M}, where
 all the $u$, $d$ and $s$ quarks have contributions. In Refs.\cite{Huang04M,Wang05M},
the  diquark-triquark type interpolating current $J(x)$ is used,
\begin{eqnarray}
J(x)={1\over \sqrt{2}} \epsilon^{abc} \left\{u^T_a(x) C\gamma_5 d_b
(x)\right\} \{ u_e (x) {\bar s}_e (x) i\gamma_5 d_c(x) - d_e (x)
{\bar s}_e (x) i\gamma_5 u_c(x)  \} \, .
\end{eqnarray}
Although the diquark-diquark-antiquark type and diquark-triquark
type configurations implemented by the interpolating currents
$\eta(x)$ and $J(x)$ respectively can give satisfactory masses for
the pentaquark state $\Theta^+(1540)$,  the resulting magnetic
moments are substantially different, once the magnetic moment can be
extracted from the electro- or photo-production experiments, we can
select the preferred  configuration. In this article, we have
neglected the contributions from the direct instantons as the
effects are supposed to be small.  In Ref.\cite{Forkel}, the authors
calculate the leading direct instanton contributions to the operator
product expansion of the nucleon correlation function with the Ioffe
current
\begin{eqnarray}
J_{p}(x)&=&\epsilon^{abc}[u^T_a(x)C\gamma_{\alpha}u_{b}(x)]\gamma_{5}
\gamma^{\alpha} d_{c}(x) \nonumber
\end{eqnarray}
in an external electromagnetic field, and find the instanton
contributions affect only the chiral odd sum rules which had
previously been considered unstable. The general form of the proton
current can be written as \cite{Baryon}
\begin{eqnarray}
J_p(x,t) &=&  \epsilon_{abc} \left\{ \left[ u_a^T(x) C d_b(x)
\right] \gamma_5 u_c(x) + t  \left[ u_a^T(x) C \gamma_5 d_b(x)
\right]  u_c(x) \right\},
\end{eqnarray}
in the limit $t=-1$, we recover the Ioffe current.
 In this article, we take the
value of the $t$ to be  $t=-1$ in Eq.(5). The pentaquark currents in
Eqs.(3-5) have the analogous Dirac structure as the baryon current
in Eq.(15), so the contributions from the direct instantons may not
affect significantly  about our analysis of the chiral even Dirac
structure in Eq.(10).
 Furthermore,
our previous  work on the pentaquark mass using the interpolating
current
\begin{eqnarray}
 J(x)&=&\epsilon^{abc}\epsilon^{def}\epsilon^{cfg}
   \{u_a^T(x)Cd_b(x)\}\{u_d^T(x)C\gamma_5 d_e(x)\}C\bar{s}_g^T(x)
   \nonumber
\end{eqnarray}
indicates  the direct instantons have neglectable contributions
\cite{WangMass}. The straightforward calculations and tedious
analysis about the direct instanton contributions to the magnetic
moment of the pentaquark state $\Theta^+(1540)$ will be our next
work.

 Here we will take a short
digression and make some discussion about the condensates and
light-cone amplitudes in Eqs.(12-13). The presence of the external
electromagnetic field $F_{\mu\nu}$ induces three new vacuum
condensates i.e. the vacuum susceptibilities in the QCD vacuum
\cite{Ioffe84},
\begin{eqnarray}
\langle  {\overline  q} \sigma_{\mu\nu} q  \rangle_{F_{\mu\nu}} =
e_q  \chi
F_{\mu\nu} \langle  {\overline  q}  q  \rangle \, , \nonumber \\
g_s \langle  \bar{q} G_{\mu\nu} q  \rangle_{F_{\mu\nu}} = e_q
\kappa F_{\mu\nu}
\langle  {\overline  q}  q  \rangle \, , \nonumber\\
g_s \epsilon^{\mu\nu\lambda\sigma} \langle  {\overline  q} \gamma_5
G_{\lambda\sigma} q  \rangle_{F_{\mu\nu}} = i e_q  \xi F^{\mu\nu}
\langle  {\overline  q}  q  \rangle \, , \nonumber
\end{eqnarray}
where $e_q$ is the quark charge, the $\chi$, $\kappa$ and $\xi$  are
the quark vacuum susceptibilities. The values with  different
theoretical approaches are different from each other, for a short
review,  one can see Ref.\cite{Wang02}. Here we shall adopt the
values $\chi=-4.4\, \mbox{GeV}^{-2}$, $\kappa =0.4$ and $\xi = -0.8$
\cite{Ioffe84,LightCone,Belyaev84}. In calculation,
 we have neglected the terms which concern the $ g_s \langle  \bar{q} G_{\mu\nu} q  \rangle$ and
 $g_s \epsilon^{\mu\nu\lambda\sigma}
\langle  {\overline  q} \gamma_5  G_{\lambda\sigma} q |0 \rangle$
induced   vacuum susceptibilities  as they are suppressed by large
denominators.  The photons can couple to the quark lines
perturbatively and nonperturbatively,
  which results in  two classes of diagrams. In the first class of diagrams,
  the photons couple to
the quark lines perturbatively through the standard QED, the second
class of diagrams involve the nonperturbative interactions of
photons with the quark lines, which are parameterized by  the photon
light-cone distribution amplitudes instead of the vacuum
susceptibilities.  In this article, the following two-particle
photon light-cone distribution amplitude has contributions to the
magnetic moment \cite{LightCone,photon},
\begin{eqnarray}
\langle \gamma (q)|\bar q(x) \gamma_\mu\gamma_5q(0)|0\rangle& =&
\frac{f}{4} e_q \epsilon_{\mu\nu\rho\sigma} \varepsilon^\nu q^\rho
x^\sigma \int_0^1 due^{iuqx} \psi(u) \, ,
\end{eqnarray}
where the $\psi(u)$ is the  twist--2 photon light-cone distribution
amplitudes.

We make  Borel transform  with respect to the variable $P^2$ in
Eq.(12) and double Borel transform  with respect to the variables
$p^2$ and $(p+q)^2$ in Eq.(13),
 \begin{eqnarray}
\Pi (M^2) \equiv \lim_{n,P^2 \rightarrow \infty} \frac{1}{\Gamma(n)}
(P^2)^{n} \left( - \frac{d}{d P^2} \right)^n \Pi (P^2), \\
{{\cal B}}^{M_1^2}_{(p+q)^2} {{\cal  B}}^{M_2^2}_{p^2} {\Gamma
(n)\over [ m^2 -(1-u)(p+q)^2-up^2]^n }= (M^2)^{2-n} e^{-{m^2\over
M^2}} \delta (u-u_0 ) ,
\end{eqnarray}
  with  $M^2 = P^2/n$ in Eq.(17) and
  $M^2=\frac{M_1^2M_2^2}{M_1^2+M_2^2}$,
$u_0\equiv\frac{M_1^2}{M_1^2+M_2^2}$ in Eq.(18). Finally we obtain
the sum rules,
\begin{eqnarray}
 \left\{F_1(0)+F_2(0)\right\}
(1+CM^2)f_0^2e^{-\frac{m^2_{\Theta^+}}{M^2}}&=&-4e_s AA, \\
\{ F_1(0)+F_2(0)\}f_0^2 e^{-\frac{m_{\Theta^+}^2}{M^2}}&=&-e_s BB,
\end{eqnarray}
where
\begin{eqnarray}
AA &=& -\frac{(5t^2+2t+5)M^{12}E_4}{2^{14}5!\pi^8}
+\frac{(5t^2+2t+5)m_s\chi \langle \bar{s}s\rangle
M^{10}E_3}{2^{11}5!\pi^6} \nonumber \\
&&+\frac{(5+2t-7t^2)\langle \bar{q}q\rangle^2M^6E_1}{2^{9}3^2\pi^4}
-\frac{(5+2t-7t^2)m_s \langle \bar{q}q\rangle^2 \langle
\bar{s}s\rangle \chi
M^4 E_0}{2^{7}3^2\pi^2}\nonumber\\
&&-\frac{(5t^2+2t+5)M^8E_2}{2^{16}4!\pi^6}\langle \frac{\alpha_s
GG}{\pi} \rangle  +\frac{(5t^2+2t+5)\langle
\bar{q}q\rangle^4}{2^{5}3^3 },  \nonumber\\
BB
&=&\frac{(5t^2+2t+5)M^{12}E_4}{2^{12}5!\pi^8}-\frac{(5t^2+2t+5)M^{10}E_3f
\psi(u_0)}{2^{15}5\pi^6}\nonumber\\
&&+\frac{(7t^2-2t-5)\langle
\bar{q}q\rangle^2M^6E_1}{2^{7}3^2\pi^4}-\frac{(7t^2-2t-5)\langle
\bar{q}q\rangle^2f
\psi(u_0)M^4E_0}{2^{6}3\pi^2}\nonumber \\
&&+\frac{(5t^2+2t+5)M^8E_2}{2^{14}4!\pi^6}\langle \frac{\alpha_s
GG}{\pi} \rangle  -\frac{(5t^2+2t+5)f
\psi(u_0)M^6E_1}{2^{14}3\pi^4}\langle \frac{\alpha_s GG}{\pi}
\rangle \nonumber \\
&&-\frac{(5t^2+2t+5)\langle \bar{q}q\rangle^4}{2^{3}3^3 },
\nonumber \\
E_n&=&1-exp\left(-\frac{s_0}{M^2}\right)\sum_{k=0}^{n}\left(\frac{s_0}{M^2}\right)^k\frac{1}{k!}\,
. \nonumber
\end{eqnarray}

The Borel transform in Eq.(17) can not eliminate the contaminations
to the correlation function from the single-pole terms, we introduce
 the parameter  $C$ which proportional to the  $C_{subtract}$ in Eq.(11) to the subtract the contaminations.
  We have no knowledge about the  electromagnetic transitions between the pentaquark state $\Theta^+(1540)$
 and the excited states (or high resonances), the $C$ can be taken as a free parameter, we  choose the suitable values for $C$ to
  eliminate the contaminations from  the single-pole terms to obtain the reliable  sum rules. The contributions from
  the single-pole terms may as large as or larger than the double-pole term, in practical calculations,
   the $C$ can be fitted to give stable sum rules with respect to
 variations  of the Borel parameter $M^2$ in a suitable interval. Taking  the $C$ as an  unknown constant has
 smeared the complex energy $s$ and high resonances masses dependence, which will certainly impair the
 prediction
 power. As final numerical results are insensitive to the
 threshold parameter $s_0$ and there realy exists a platform with the variations  of the Borel parameter $M^2$, the
 predictions still make sense.
   The double Borel transform in Eq.(18) can eliminate
  the single-pole terms naturally; for more discussions
about the double Borel transform, one can consult
Ref.\cite{Belyaev94}. Furthermore, from the correlation function
$\Pi_0(p)$ in Eq.(1), we can obtain the sum rules for the coupling
constant $f_0$ \cite{Matheus04},
\begin{equation}
f^2_0e^{-{m_{\Theta^+}^2\over M^2}}=CC,
\end{equation}

\begin{eqnarray}
CC &=&\frac{3(5t^2+2t+5)M^{12}E_5}{ 2^{11}7! \pi^8}
+\frac{(5t^2+2t+5)m_s\langle\bar{s}s\rangle M^{8}E_3}{2^{10}5!
\pi^6}+ \frac{(1-t)^2
M^{8}E_3}{2^{13}5!\pi^6}\langle \frac{\alpha_s GG}{\pi} \rangle \nonumber\\
& &+  \frac{(7t^2-2t-5)\langle \bar{q}q\rangle^2
M^{6}E_2}{2^{9}3^2\pi^4}- \frac{(5t^2+2t+5) m_s \langle
\bar{s}g_s\sigma  G s\rangle  M^{6}E_2}{2^{14}3^2\pi^6}
\nonumber\\
&  &+\frac{(7t^2-2t-5) m_s \langle\bar{s}s\rangle \langle \bar{q}
q\rangle^2M^{2}E_0}{2^{6}3^2\pi^2} +\frac{(5t^2+2t+5)\langle \bar{q}
q\rangle^4}{6^3}\, .\nonumber
\end{eqnarray}
 From above equations, we can obtain the sum rules for the form
 factor $F_1(0)+F_2(0)$,
\begin{eqnarray}
 \left\{F_1(0)+F_2(0)\right\}\left\{1+CM^2\right\} =-4 e_s \frac{
 AA} {CC} \,  , \\
 \left\{F_1(0)+F_2(0)\right\} =- e_s \frac{
 BB} {CC} \,  .
\end{eqnarray}

\section{Numerical Results and discussions}
In this article, we take the value of the parameter $t$ for the
interpolating current $\eta(x)$  to be $t=-1$, which can give stable
mass for the pentaquark state $\Theta^+(1540)$ (i.e.
$m_{\Theta^+}\approx 1540~MeV$) with respect to the variations of
the Borel mass $M^2$ in the considered interval $M^2=(2-3)GeV^2$
\cite{Matheus04}. The parameters for the condensates are chosen to
be the standard values, although there are some suggestions for
updating those values, for reviews, one can see Ref.\cite{Update},
$\langle \bar{s}s \rangle=(0.8\pm0.1) \langle \bar{u}u \rangle$,
$\langle \bar{q}q \rangle=\langle \bar{u}u \rangle=\langle \bar{d}d
\rangle=-(240\pm 10 MeV)^3$, $\langle \bar{s}g_s\sigma G s
\rangle=m_0^2\langle \bar{s}s \rangle$, $m_0^2=(0.8\pm0.1)GeV^2$,
$\chi=-(4.4\pm 0.4)GeV^{-2}$, $\langle \frac{\alpha_s GG}{\pi}
\rangle=(0.33GeV)^4$,
 $m_u=m_d=0$ and $m_s=(140\pm10) MeV$. Small variations of those condensates will not
 lead to large  changes about   the numerical
 values.  The threshold parameter $s_0$ is chosen to  vary between $(3.6-4.4) GeV^2$ to avoid possible contaminations from
 the high resonances and continuum states, which is shown in Fig.1 for $\langle \bar{s}s \rangle=0.8
   \langle \bar{u}u \rangle$,
$\langle \bar{q}q \rangle=-(240 MeV)^3$,  $m_0^2=0.8GeV^2$,
$\chi=-4.4GeV^{-2}$  and $m_s=140 MeV$. In the EFSR,  for $s_0=4.0
GeV^2$ and $M^2=(2-5)GeV^2$, we obtain the values
 \begin{eqnarray}
 F_1(0)+F_2(0)&=&-(0.18\pm 0.03) \, , \nonumber\\
 \mu_{\Theta^+}&=& -(0.18\pm 0.03) \frac{e_{\Theta^+}}{2m_{\Theta^+}}\, , \nonumber \\
  &=& -(0.11\pm 0.02)\mu_N,
  \end{eqnarray}
where the $\mu_N$ is the nucleon  magneton.
 From the Table 1,  we can see that although  the numerical values for
 the magnetic moment of the pentaquark  state $\Theta^+(1540)$  vary with theoretical
 approaches, they are small in general;  our numerical results
are consistent with most of the existing values of theoretical
estimations in magnitude, however, with negative sign. In Eq.(22),
the perturbative contributions are about $25\%$, the dominating
contributions come from the dimension-6 quark condensates terms
$\langle\bar{q}q\rangle^2$ , the contributions from the gluon
condensate $\langle \frac{\alpha_s}{\pi}GG\rangle$, dimension-12
quark condensate $\langle\bar{q}q\rangle^4$ are very small and can
be safely neglected. In calculation, other vacuum condensates are
neglected due to the suppression of the large denominators, the
truncation of the operator product expansion makes  sense.

For the conventional ground state mesons and baryons, due to the
resonance dominates over the QCD continuum contributions, the good
convergence of the operator product expansion, and the useful
experimental guidance on the threshold parameter $s_0$, we can
obtain the fiducial Borel mass region.  However, in the QCD sum
rules for the pentaquark states, the spectral density $\rho(s) \sim
s^m$ with $m$ larger than  the corresponding ones in the
 sum rules for the conventional baryons, larger $m$ means   stronger dependence on the
continuum or the threshold parameter $s_0$ \cite{Narison04,Zhu05}.
In Eq.(21), due to the huge continuum contributions, the  predicted
mass increases with the continuum threshold $s_0$, the QCD sum rules
cannot strictly indicate the existence of the resonance in the
spectral function,  the threshold parameter  $s_0$ has  to be fixed
ad hoc or intuitively. In  the finite energy sum rules (FESR)
approach, the exponential weight function $exp(-\frac{s}{M^2})$ is
replaced by $s^n$ in the numerical analysis. The FESR   correlate
the ground state mass and the QCD continuum threshold $s_0$, and
separate the ground state and QCD continuum contributions from the
very beginning, for  some pentaquark currents, there happen exist
reasonable stability regions  $s_0$ \cite{Narison04}. The weight
function $s^n$ enhances the continuum or the high mass resonances
rather than the lowest ground state, we must make sure that only the
lowest pole terms contribute to the FESR below the  $s_0$, in some
case, a naive stability region  $s_0$ can not guarantee    a
physically reasonable value for the $s_0$ \cite{Zhu05}. It is
obviously, as QCD models, the Laplace sum rules and the FESR have
both  advantages and shortcomings. Although the quantities $f^2_0
exp(-{m_{\Theta^+}^2\over M^2})$ in Eq.(19) and Eq.(21) have strong
dependence on the continuum threshold parameter $s_0$, the
dependence is eliminated in Eq.(22) which results in a net $s_0$
insensitivity;  it is far from the ideal case, where the $f^2_0
exp(-{m_{\Theta^+}^2\over M^2})$ are insensitive to the threshold
parameter $s_0$. From Fig.2, we can see that for the Borel parameter
$M^2=(2-5)GeV^2$, the contributions of the double-pole and
single-pole terms below the threshold $s_0=4.0GeV^2$ are dominating,
for example, the contributions of the continuum from the
$s_0=4.0GeV^2$ to a large interval $s_0=20.0GeV^2$ will not exceed
$30\%$ (comparing with the contributions below $s_0=4.0GeV^2$), the
lowest pole terms dominance still hold and we have the fiducial
Borel mass domain where the neglect of higher-order terms in the
short-distance expansion is justified while the nucleon pole terms
still dominate  over the continuum.

In the EFSR, the Borel transform can not eliminate the contributions
from the single-pole terms, on the other hand, we have no knowledge
about the transitions between the ground states $\Theta^+(1540)$
state and excited states (or high resonances), in practical
manipulations, we can introduce some free parameters which denote
the contributions from the single-pole terms and subtract them. In
choosing the parameter $C$ in Eq.(22), we must take care, in
general, the $C$ can be chosen to give the stable sum rules  with
respect to the variations of the Borel parameter $M^2$. Although the
uncertainty of the condensates, the neglect of the higher dimension
condensates, the lack of perturbative QCD corrections, etc, will
result in errors, we have stable sum rules for the magnetic moment,
furthermore, small variations of the condensates will not result in
large changes for the values,  the predictions are qualitative at
least. If the pentaquark state $\Theta^+(1540)$ can be interpolated
  by the linear superposition of the S-type and P-type diquark-diquark-antiquark current, $\eta(x)$,
  the predictions are quantitative and make sense.
Comparing with Ref.\cite{Wang05M}, the EFSR  with different
interpolating currents (standing for different configurations) can
lead to very different results for the magnetic moment, although
they can both give satisfactory masses for the pentaquark state
 $\Theta^+(1540)$.

 In the LCSR, the double Borel transform
is supposed to  eliminate the single-pole terms which concerning the
contaminations from high resonances  and continuum states, and gives
stable sum rules  with the variations of the Borel parameter $M^2$.
However, that is not the case, from Fig.3, we can see that the LCSRs
for the form factor $-[F_1(0)+F_2(0)]$ are sensitive to the
threshold parameter $s_0$, when $s_0>4.0 GeV^2$, the values of the
form factor $-[F_1(0)+F_2(0)]$ decrease drastically with the
increase of the Borel parameter $M^2$ and there will not appear the
platform. For $s_0=3.8GeV^2$, $M^2=(2-5)GeV^2$, $ \psi(u)=1$ and
$f=0.028 \mbox{GeV}^2$ \cite{LightCone,photon},
\begin{eqnarray}
 F_1(0)+F_2(0)&=&-(0.2-0.8) \, , \nonumber\\
 \mu_{\Theta^+}&=& -(0.2-0.8) \frac{e_{\Theta^+}}{2m_{\Theta^+}}\, , \nonumber \\
  &\approx& -(0.1-0.5)\mu_N.
  \end{eqnarray}
In above equations, we take the maximal variation  interval for the
values of the form factor $F_1(0)+F_2(0)$.

\begin{figure}
 \centering
 \includegraphics[totalheight=7cm]{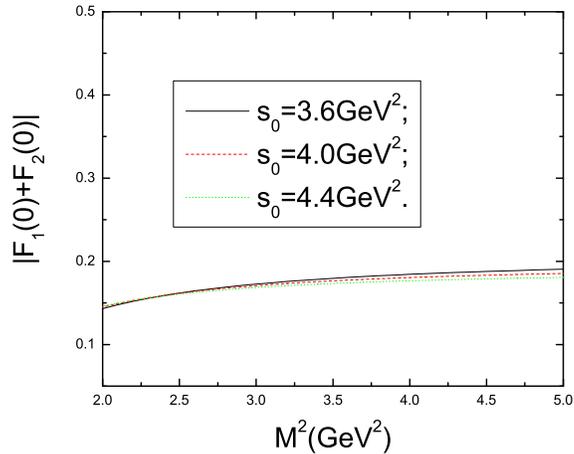}
 \caption{$|F_1(0)+F_2(0)|$ with the variations of the Borel Parameter $M^2$ for EFSR. }
\end{figure}

\begin{figure}
 \centering
 \includegraphics[totalheight=7cm]{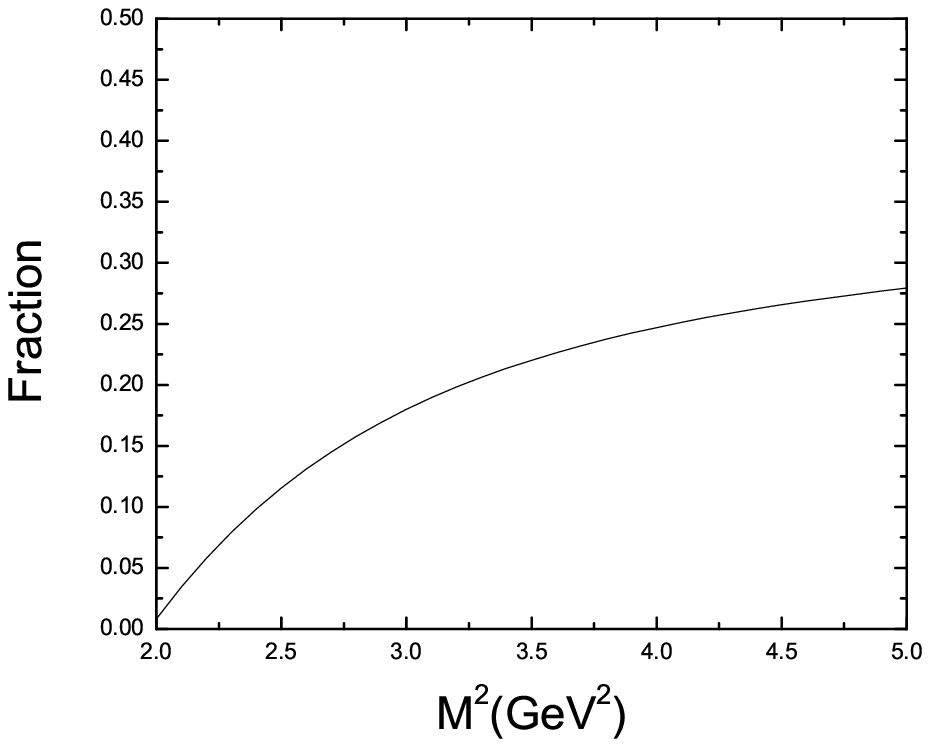}
 \caption{$|\frac{[\frac{AA}{CC}]_{s0=20.0GeV^2}-[\frac{AA}{CC}]_{s0=4.0GeV^2}}{[\frac{AA}{CC}]_{s0=4.0GeV^2}}|$ with the variations of the Borel Parameter $M^2$ for EFSR. }
\end{figure}

In the LCSR approach, the uncertainty of the photon light-cone
distribution amplitudes, keeping only the lowest-twist few terms of
two particles distribution amplitudes, the uncertainty of the
condensates, the neglect of the higher dimension condensates, the
lack of perturbative QCD corrections, etc, can lead to errors in the
predictions. The LCSRs for the form factor $F_1(0)+F_2(0)$ ( see
Eq.(23) and Eq.(25) ) are  very sensitive to the values of the
parameter $f$, small variations of the $f$ can lead to large changes
for the magnetic moment, which is shown in Fig.4. The inclusion of
the contributions from the direct instantons may improve the
stability of the LCSR, to our knowledge, there exist no such type
works on the magnetic moments of the baryon with the LCSR, this may
be our next work.  In this article, only the nonperturbative
interactions of the photons with the $s$ quark line of the form
$\langle \gamma (q)|\bar q(x) \gamma_\mu\gamma_5q(0)|0\rangle$ have
contributions to the magnetic moment which is different
significantly from the corresponding ones with the interpolating
current $J(x)$ in Eq.(14), where all the non-local matrix elements
$\langle \gamma (q)|\bar q(x) \sigma_{\mu\nu}q(0)|0\rangle$,
$\langle \gamma (q)|\bar q(x) \gamma_\mu\gamma_5q(0)|0\rangle$, $
\langle \gamma (q)|\bar q(x) \gamma_\mu q(0)|0\rangle$ have
contributions to the magnetic moment, and result in  more stable sum
rules \cite{Huang04M}. In the vector dominance model,
\begin{eqnarray}
f \simeq \frac{f_\rho m_\rho}{g_\rho}  \simeq 0.028 GeV^2,
\end{eqnarray}
with $g_\rho = 5.5$ and $f_\rho = 0.2 GeV$.  To the leading twist
accuracy,  the light-cone amplitude $ \psi(u) $ is a constant which
is set to be unity due to the normalization condition.
\begin{figure}
 \centering
 \includegraphics[totalheight=7cm]{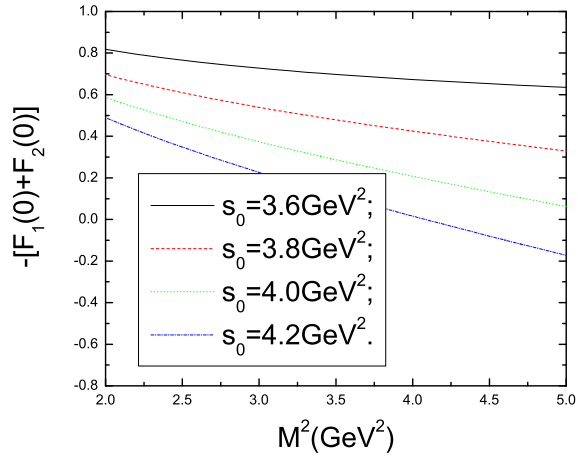}
 \caption{$-[F_1(0)+F_2(0)]$ with the variations of the Borel Parameter $M^2$ for LCSR. }
\end{figure}

\begin{figure}
 \centering
 \includegraphics[totalheight=7cm]{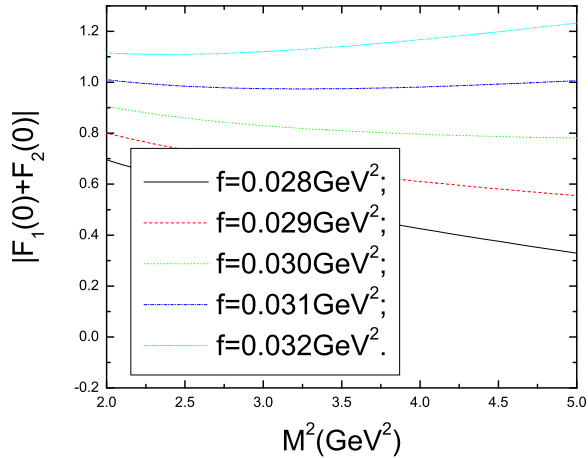}
 \caption{$|F_1(0)+F_2(0)|$ for different $f$ with the variations of the Borel Parameter $M^2$ for  LCSR ($s_0=3.8GeV^2$). }
\end{figure}

 We choose different tensor
structures for analysis in different approaches i.e. the EFSR  and
the LCSR  with the same interpolating current $\eta(x)$, the
resulting different sum rules always lead to different predictions
\cite{Huang04M,Wang05M}. The LCSRs with  the interpolating current
$\eta(x)$ depend heavily on the values of the parameter $f$ and
sensitive to the threshold parameter $s_0$,   the values obtain from
the LCSR ( see Eq.(25)) are not as reliable as the corresponding
ones from the EFSR ( see Eq.(24)).
 When the experimental measurement of the magnetic moment of the
pentaquark state $\Theta^+(1540)$
  is possible in the  future, we might  be able  to test the theoretical
   predictions,  select
the preferred quark configurations.

  \begin{table}[ht]
         \caption{The values of $\mu_{\Theta^+}$ (in unit of $\mu_N$)}
         \begin{center}
         \begin{tabular}{c||c}
         \hline\hline
         Reference          & $\mu_{\Theta^+}$ \\
                            &  $(\mu_N)$  \\ \hline\hline
   \cite{Huang04M}         &   0.12 $\pm$ 0.06        \\ \hline
         \cite{Zhao04M}     &   0.08 $\sim$ 0.6        \\ \hline
         \cite{Kim04M} & 0.2$\sim$0.3     \\ \hline
     \cite{Nam04M}        & 0.2$\sim$0.5         \\ \hline
      \cite{Zhu04M}              & 0.08 or 0.23 or 0.19 or 0.37           \\ \hline
      \cite{Inoue04M}              & 0.4          \\ \hline
               \cite{Bijker04M}             & 0.38          \\ \hline
               \cite{Goeke04M}             &-1.19 or -0.33         \\ \hline
                \cite{Hong04M}             &0.71 or 0.56          \\ \hline
      \cite{Delgado04M}            &0.362          \\ \hline

                      \cite{Wang05M} &0.24$\pm$0.02           \\ \hline\hline
                   This Work &   -(0.11$\pm$ 0.02)        \\ \hline
         \end{tabular}
         \end{center}
         \end{table}
\section{Conclusion }

In summary, we have calculated  the magnetic moment of the
 pentaquark state  $\Theta^+(1540)$ as diquark-diquark-antiquark
($[ud][ud] \bar{s}$) state with both the  EFSR and LCSR. We choose
different tensor structures for analysis in different approaches
(i.e. the EFSR and the LCSR) with the same interpolating current ,
the resulting different sum rules always lead to different
predictions \cite{Huang04M,Wang05M}. The EFSRs for the magnetic
moment are stable with the variations of the Borel parameter $M^2$
and insensitive to the threshold parameter $s_0$ (See Fig.1 and
Fig.2); the LCSRs in this work depend heavily on the values of the
non-local matrix element $\langle \gamma (q)|\bar q(x)
\gamma_\mu\gamma_5q(0)|0\rangle$, small variations of the values $f$
can result in large changes about the magnetic moments. The values
from the EFSR are more reliable than the corresponding ones from the
LCSR. The numerical results from the EFSR are consistent with most
of the existing values of theoretical estimations in magnitude,
however, with negative sign, $\mu_{\Theta^+}= -(0.18\pm 0.03)
\frac{e_{\Theta^+}}{2m_{\Theta^+}}
  = -(0.11\pm 0.02)\mu_N$ . Comparing
with Ref.\cite{Wang05M}, the EFSR with different interpolating
currents (standing for different configurations) can lead to very
different results for the magnetic moment, although they can both
give satisfactory masses for the pentaquark state
 $\Theta^+(1540)$.   If the pentaquark state
$\Theta^+(1540)$ can be interpolated
  by the linear superposition of both S-type and P-type diquark-diquark-antiquark current $\eta(x)$,
  the predictions are quantitative and make sense.
   The  magnetic moments of the
baryons are  fundamental parameters as their masses, which have
copious  information  about the underlying quark structures,
different substructures can lead to
 very different results. The width of the pentaquark state $\Theta^+(1540)$ is so
narrow, when the
 small magnetic moment can be extracted from electro- or photo-production
experiments eventually in the  future, which may be used to
distinguish the preferred configurations  from various theoretical
models, obtain more insight into the relevant degrees of freedom and
deepen our understanding about  the underlying dynamics that
determines the properties of the exotic pentaquark states.

\section*{Acknowledgment}
This  work is supported by National Natural Science Foundation,
Grant Number 10405009,  and Key Program Foundation of NCEPU. The
authors are indebted to Dr. J.He (IHEP), Dr. X.B.Huang (PKU) and Dr.
L.Li (GSCAS) for numerous help, without them, the work would not be
finished.

\end{document}